\documentclass[superscriptaddress,12pt,tightenlines,aps,nofootinbib,pre]{revtex4-1}
\usepackage{color,epsfig,graphics,amssymb,amsmath,subeqnarray,graphicx,amsthm,subfigure,mathrsfs,bm,color,xcolor,soul}

\def\Ca{\rm{Ca}}
\def\x{{\mathbf{x}}}

\def\u{{\mathbf{u}}}
\def\M{{\mathbf{M}}}

\def\q{{\mathbf{q}}}
\def\f{{\mathbf{f}}}
\def\n{{\mathbf{n}}}

\begin{document}
\title{Edge states control droplet break-up in sub-critical extensional flows}
\author{Giacomo Gallino}
\email{giacomo.gallino@epfl.ch}
\affiliation{Laboratory of Fluid Mechanics and Instabilities, \'{E}cole Polytechnique F\'{e}d\'{e}rale de Lausanne. EPFL-STI-IGM-LFMI, CH-1015 Lausanne, Switzerland}

\author{Tobias M. Schneider}
\email{tobias.schneider@epfl.ch}
\affiliation{Emergent Complexity in Physical Systems Laboratory, \'{E}cole Polytechnique F\'{e}d\'{e}rale de Lausanne. EPFL-STI-IGM-ECPS, CH-1015 Lausanne, Switzerland}

\author{Fran\c{c}ois Gallaire}
\email{francois.gallaire@epfl.ch}
\affiliation{Laboratory of Fluid Mechanics and Instabilities, \'{E}cole Polytechnique F\'{e}d\'{e}rale de Lausanne. EPFL-STI-IGM-LFMI, CH-1015 Lausanne, Switzerland}

\date{\today}

\begin{abstract}
A fluid droplet suspended in an extensional flow of moderate intensity may break into pieces, depending on the amplitude of the initial droplet deformation. In subcritical uniaxial extensional flow the non-breaking base state is linearly stable, implying that only a finite amplitude perturbation can trigger break-up. Consequently, the stable base solution is surrounded by its finite basin of attraction. The basin boundary, which separates initial droplet shapes returning to the non-breaking base state from those becoming unstable and breaking up, is characterized using edge tracking techniques. We numerically construct the edge state, a dynamically unstable equilibrium whose stable manifold forms the basin boundary. The edge state equilibrium controls if the droplet breaks and selects a unique path towards break-up. This path physically corresponds to the well-known end-pinching mechanism. Our results thereby rationalize the dynamics observed experimentally [Stone \& Leal, J. Fluid Mech. 206, 223 (1989)].

\end{abstract}

\maketitle

\section{Introduction}

The shape of a droplet in low Reynolds number flows results from two competing effects, the viscous forces stretching the droplet, and the surface tension forces reducing its deformation in order to minimize the droplet surface area. When the ratio between viscous forces and surface tension forces, defined as the capillary number $\Ca$, is large, the droplet may deform until breaking into smaller droplets~\cite{acrivos1983breakup,rallison1984deformation,stone1994dynamics}.

G. I. Taylor was the first to systematically study this phenomenon, by placing a liquid drop in flow fields generated by counter-rotating rollers~\cite{taylor1934formation}. Imposing a hyperbolic flow, he observed that when the capillary number exceeds a critical value $\Ca_\text{crit}$, the droplet always breaks. Consistent results have later been obtained in experimental~\cite{rumscheidt1961particle,bentley1986experimental} and numerical~\cite{rallison1981numerical} studies. At the critical capillary number, the non-breaking equilibrium state, or {\it base state}, undergoes a saddle-node bifurcation~\cite{barthes1973deformation,blawzdziewicz2002critical} and  no longer exists for $\Ca > \Ca_\text{crit}$.  
Interestingly, the droplet can break-up even for a subcritical value of Ca, depending on its initial shape~\cite{stone1986experimental,stone1989relaxation,stone1989influence}. The dependence of the droplet stability upon the initial shape indicates the existence of a finite basin of attraction surrounding the base state.  Initial droplet shapes evolving towards the base state are separated from those breaking apart by the basin boundary.

Due to the high dimensionality of the state space and the infinite number of possible initial shapes, it is challenging to characterize the basin of attraction and predict if a droplet breaks up. The situation is however analogous to other high-dimensional nonlinear dynamical systems characterized by a finite basin of attraction, as for instance in pipe flow and plane Couette flow, where the laminar base flow is linearly stable and a finite amplitude perturbation is needed to trigger turbulence.  In the latter cases, recent studies have demonstrated the relevance of unstable equilibrium states embedded in the basin boundary. Of specific importance are {\it edge states}~\cite{skufca2006edge,schneider2007turbulence}, unstable equilibria in the basin boundary with only a single unstable direction. These edge states are attracting for the dynamics confined to the basin boundary but unstable in the direction perpendicular to the boundary. Their significance lies in their guiding role for the transition between laminar and turbulent flows.

In this paper we show that an unstable edge state equilibrium embedded in the basin boundary of the base state governs the break-up dynamics of a droplet in a sub-critical uniaxial extensional flow. In fact, the unstable direction of the edge state selects an almost unique path towards droplet break-up.

\section{Numerical method}

\begin{figure}

\center

\includegraphics[width=5cm]{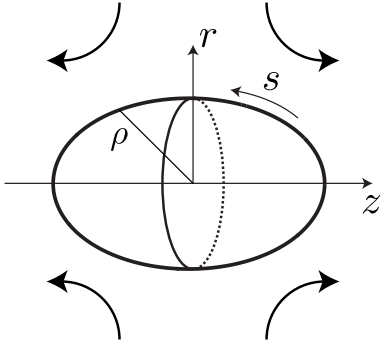}

\caption{Sketch: the axisymmetric droplet, its parametrization and the coordinate system. The arrows indicate the direction of the extensional flow.}

\label{fig:sketch}

\end{figure}

We consider a droplet of fluid $1$ and unperturbed radius $R$ suspended in fluid $2$, the viscosity ratio between the two fluids is $\lambda=\mu_1/\mu_2$ and the surface tension $\gamma$. The characteristic length scale of the problem is $R$ and the velocity scale $\gamma/\mu_2$. After nondimensionalization, the axisymmetric droplet shape is expressed in cylindrical coordinates as $\x=\left[z(t,s),r(t,s) \right]$, with $z$ the axial and $r$ the radial coordinates, $t$ the time and  $s \in [0,1]$ the spatial curvilinear coordinate along the droplet's meridian (see figure~\ref{fig:sketch}).

The unknown interface velocity $\u=\left[u_z(t,s),u_r(t,s)\right]$ on a point of the interface $\x_0$ is determined by solving the Stokes equations in fluid $1$ and $2$. The creeping flow equations can be recast into a boundary integral equation~\cite{rallison1984deformation,pozrikidis1992boundary}. Consequently, the interface velocity is given by an integration along the droplet interface of arc-length $l$

\begin{align}\label{eq:space}
4\pi (1+\lambda)\u(\x_0) = 8\pi \u_\infty-\int_0^{l} \M(\x_0,\x) \cdot \Delta \f(\x) dl (\x) +(1-\lambda) \int_0^{l} \u(\x) \cdot \q(\x_0,\x) \cdot \n(\x) dl (\x),
\end{align}
where $\M$ and $\q$ are the Green's functions of the Stokes equations forced by a ring of point forces after azimuthal integration (see Ref.~\cite{pozrikidis1992boundary} for details), $\n$ is the normal vector pointing into the suspending fluid and $\Delta \f(\x) = (\nabla_s \cdot \n)\n$ is the discontinuity in normal stresses scaled by $\gamma/R$. In the far-field we impose vanishing pressure $p \rightarrow 0$ and the uniaxial extensional flow  $\u_\infty = \Gamma \cdot \x_0$ where, in cylindrical coordinates $(z,r)$ and omitting the $\theta$ direction,
\begin{equation*}
\Gamma = \frac{\Ca}{2}
\begin{pmatrix}
2 & 0 \\
0 & -1
\end{pmatrix}.
\end{equation*}
The capillary number $\Ca=\mu_2 G R/\gamma$ and $G$ are respectively the nondimensional and dimensional velocity gradients. The droplet interface evolves in time under the impermeability condition
\begin{equation}
\frac{\text{d} \x}{\text{d}t}=\left((\u-\u_\text{drop}\right) \cdot \n)\n, \label{eq:time}
\end{equation}
where $\u_\text{drop}$ is the velocity of the center of mass of the droplet. Focussing on shape deformations, we describe the system in the frame of reference comoving with the droplet, thus enforcing  $z$-translational symmetry~\cite{rallison1984deformation}.

Equation~\eqref{eq:space} is discretized using a pseudospectral scheme in which the interface coordinates $\x(s)$, velocity $\u(s)$ and normal stresses $\Delta \f(s)$ are represented by Legendre polynomials. For direct numerical simulations (DNS), we integrate equation~\eqref{eq:time} with a second order Runge-Kutta scheme. Typically, $100$ modes are sufficient for the spatial discretization and the time step is $\Delta t=2 \times 10^{-3}$. Similar to previous studies~\cite{eggers2009numerical,zhao2013shape}, we develop a Newton solver in order to find the roots of equation~\eqref{eq:time}, which allows for the computation of unstable steady states and rigorous linear stability analysis. The code has been validated against Refs.~\cite{stone1989relaxation,koh1989stability,pozrikidis1990instability,FLM:10044765}.

\section{Results}

\subsection{Edge tracking}

\begin{figure}

\center

\includegraphics[width=10cm]{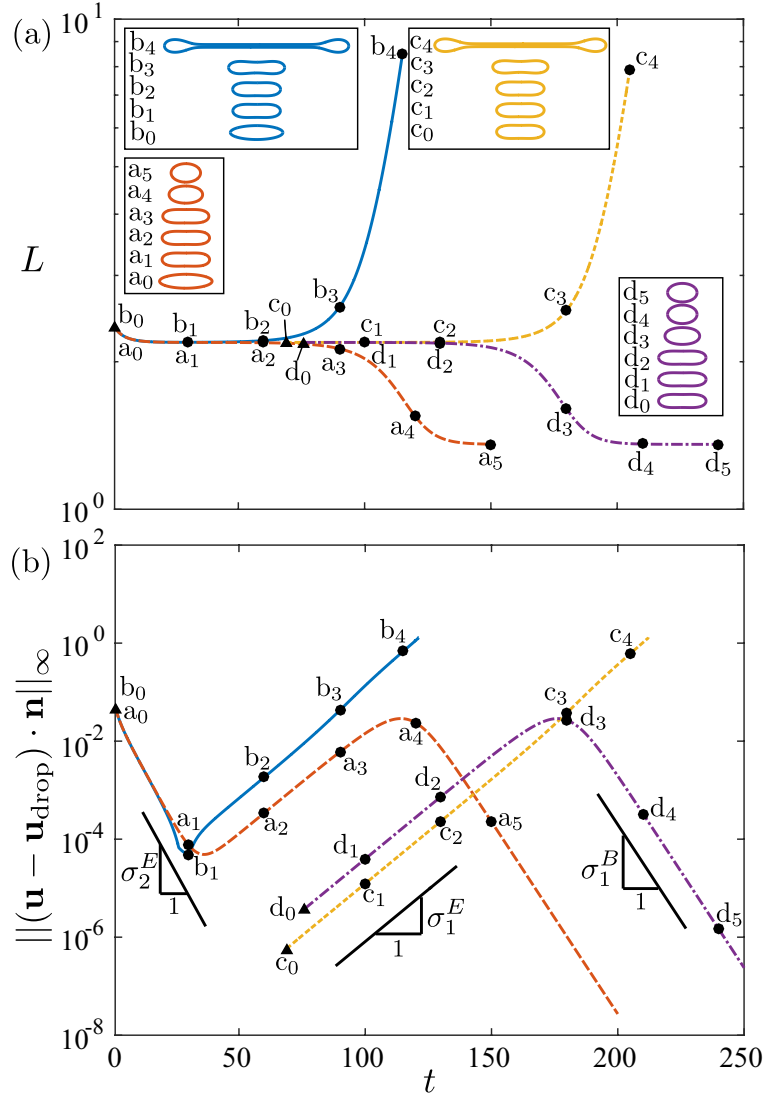}

\caption{Edge tracking orbits for $\Ca=0.1$ and $\lambda=1$. (a) Droplet elongation versus time, the orbits reach the stable base state or break-up through the end-pinching mechanism. Triangles denote the initial shapes and dots subsequent snapshots. (b) Corresponding normal velocity residuals measured in the $L_\infty$ norm versus time, the growth and decay rates are given by the most unstable and least stable eigenvalues of the edge state and the base state $\sigma_1^E$, $\sigma_2^E$ and $\sigma_1^B$ respectively.}

\label{fig:edgeTracking}

\end{figure}

To follow orbits in the basin boundary and compute the edge state we adapt the edge tracking techniques used in shear flows~\cite{schneider2007turbulence} and other fields~\cite{kreilos2017fully,virot2017stability}:
We consider two slightly different initially ellipsoidal droplets for $\Ca=0.1$ and $\lambda=1$.  One labeled $a_0$ in figure~\ref{fig:edgeTracking}a
 approaches the base state and another one, labeled $b_0$, eventually breaks apart. Consequently, the two droplets, both of identical volume, define two initial conditions on either side of the basin boundary. For fixed volume, the shape of an ellipsoidal droplet is uniquely defined by the droplet half-length $L$, so that iterative bisecting in $L$ allows to construct a pair of arbitrarily close initial conditions on opposite sides of the basin boundary. Orbits starting from those initial conditions bracket and approximate an  \emph{edge orbit} that neither returns to the base state nor evolves towards break-up but remains in the basin boundary. When the distance between the bracketing orbits becomes larger than a set threshold (usually $10^{-4}$ measured in the difference of surface area), the approximation of the edge orbit is refined and a new initial condition is created by bisecting between the current shapes~\footnote{Technically, a convex combination followed by resizing to enforce volume conservation is used to interpolate between the shapes along the two previous orbits. Iterative bisection in the weight parameter of the convex combination yields a new initial condition for bracketing the orbit.}. Iterating this procedure allows us to numerically follow an edge orbit in the basin boundary for  arbitrary time. Two iterations are reported in figure~\ref{fig:edgeTracking} with initial conditions $c_0$ (long-dashed) and $d_0$ (dashed-dotted).
 
After short time the edge orbit settles to a shape of constant $L$, indicating that a locally attracting equilibrium state in the basin boundary, the edge state, has been reached. We have verified the existence of the nonlinear edge state equilibrium by Newton iteration, reaching convergence to machine precision in a few iterations.
 The bracketing orbits transiently approach the edge state (shape a$_1$, b$_1$, c$_0$ and d$_0$), as evidenced by the low values of the residuals shown in figure~\ref{fig:edgeTracking}b, before evolving towards break-up or approaching the base state. As expected, the growth rate of the residuals shows an exponential behavior close to the edge state and to the base state. For both bracketing orbits, the least stable eigenvalue $\sigma_2^E$ of the edge state governs the attractive dynamics, while its unstable eigenvalue $\sigma_1^E$ drives the dynamics when departing from it. Likewise, when the droplet approaches the stable base state, the decay of the residuals is dictated by its least stable eigenvalue $\sigma_1^B$. It is noteworthy that the exponential growth (or decay) is maintained also far from the equilibrium states.

\begin{figure}

\center

\includegraphics[width=16cm]{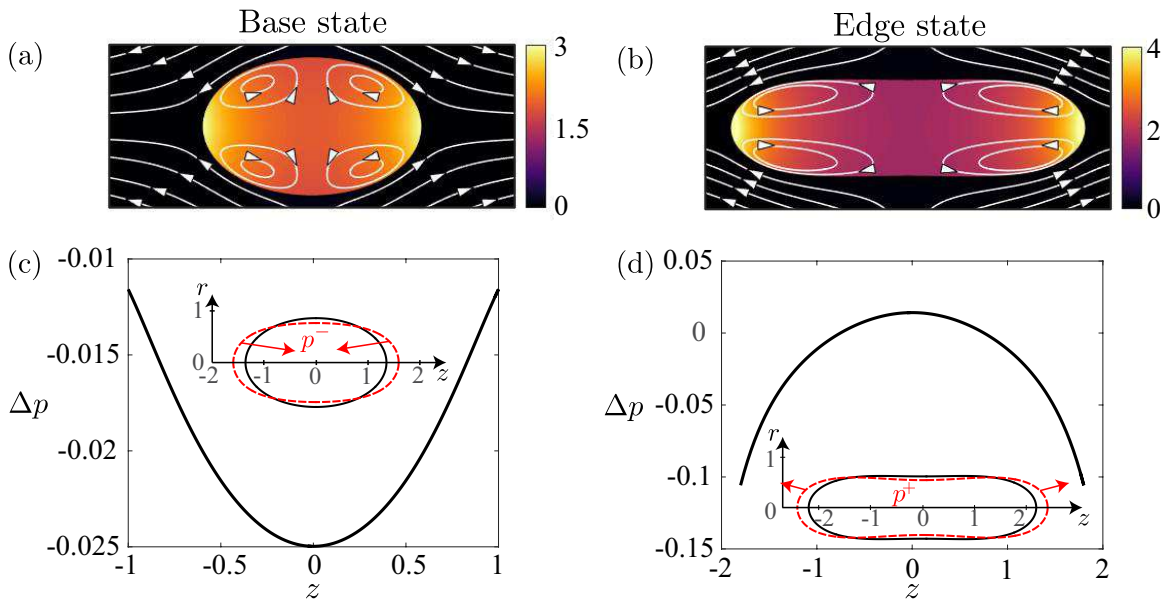}

\caption{Pressure field and velocity streamlines for the base state (a) and the edge state (b) for $\Ca=0.1$ and $\lambda=1$. (c-d) Change in pressure along the $z-$axis when the base state is perturbed with its least stable eigenmode (c) and the edge state with its most unstable eigenmode (d), the insets show the shape of the base state and the edge state (solid line) and their shapes after the modes are superimposed (dashed line). The interface motion is depicted by the arrows.}

\label{fig:physics}

\end{figure}

The flow fields associated to the equilibrium states are qualitatively similar to each other (see figure~\ref{fig:physics}a-b). Namely, the external flow induces a fluid motion along the interface toward the droplet caps, which is compensated by a recirculation along the axis driven by a pressure gradient (decreasing from the caps toward the droplet center). The equilibrium states exist when the flow along the interface is equal to the recirculating one, although they might be stable or unstable. If a slight droplet elongation increases/decreases the recirculating flow, the droplet is stable/unstable.

When the base state is perturbed with its least stable eigenmode, the pressure along the axis decreases more in the center than at the caps (see figure~\ref{fig:physics}c), therefore the recirculation becomes stronger due to the lower pressure in the droplet center, producing an interface displacement that restores the base state shape. When the edge state is perturbed with its most unstable eigenmode, the pressure along the axis increases in the center and decreases at the caps (see figure~\ref{fig:physics}d), therefore the recirculation becomes weaker due to the higher pressure in the droplet center, producing an unstable droplet elongation that leads to break-up.

\begin{figure}

\center

\includegraphics[width=10cm]{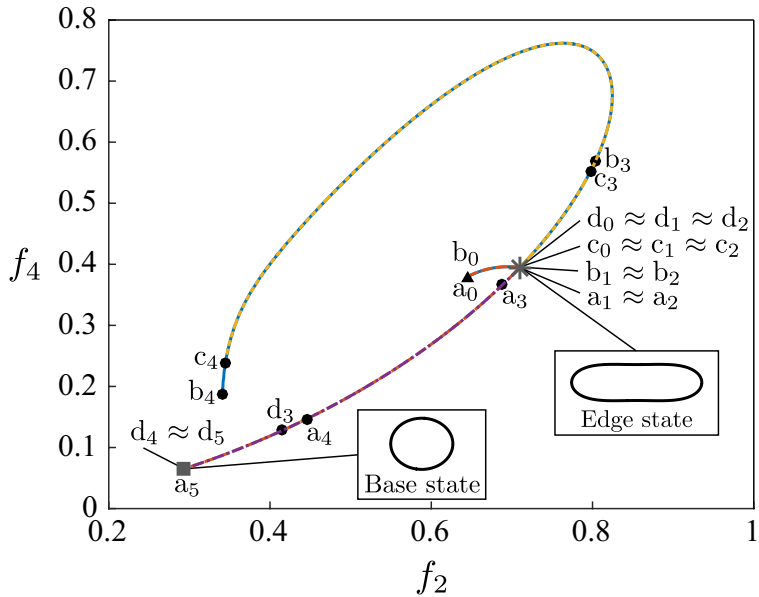}

\caption{Edge tracking orbits in state space $(f_2,f_4)$ for the cases shown in figure~\ref{fig:edgeTracking}, where $f_2$ and $f_4$ are the droplet radius projections onto the second and fourth Legendre polynomials. The star and the square symbols indicate the locations of the edge state and stable base state respectively.}

\label{fig:phaseSpace}

\end{figure}

To demonstrate the dynamical relevance of the edge state, we consider a two dimensional cut of the state space, following~\cite{blawzdziewicz2002critical,hinch1980evolution}. To this end, we project the local droplet radius $\rho(s)=\sqrt{z^2+r^2}$ onto the second and fourth Legendre polynomials, obtaining the coefficients $f_2$ and $f_4$. The state-space representation of the orbits in figure \ref{fig:edgeTracking} is plotted in figure~\ref{fig:phaseSpace}.

All orbits are attracted towards the edge state along its stable manifold, which forms the basin boundary. After passing close to the edge state, the orbit leaves along the one-dimensional unstable manifold. Depending on which side of the basin boundary the initial condition is located, the orbit either approaches the base state or evolves towards break-up along an almost unique path. The state-space visualization thus shows the guiding role of the edge state and its stable manifold which controls if a droplet undergoes break-up.

\subsection{Bifurcation diagram}

\begin{figure}

\center

\includegraphics[width=10cm]{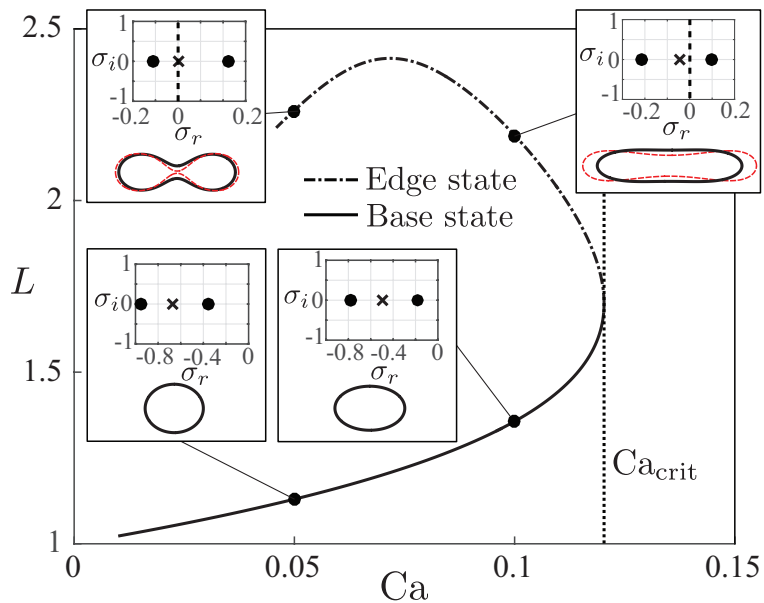}

\caption{Elongation of the droplet versus the capillary number for $\lambda=1$, the solid and dashed lines indicate the stable (base state) and unstable (edge state) branches respectively. Stable and unstable droplet shapes are plotted for $\Ca=0.05$ and $\Ca=0.1$ (solid line), superimposing the most unstable mode (dashed line). The eigenvalue spectra is also reported, circles indicate eigenvalues corresponding to symmetric modes while crosses to asymmetric modes. The translational symmetry is canceled by constraining the droplet center of mass in the origin.}

\label{fig:bifurcationDiagram}

\end{figure}

In order to track equilibrium solutions (i.e. the base state and the edge state) when varying the capillary number, we apply a continuation method based on our Newton algorithm. In particular, we implement pseudo arc-length continuation, which consists in adding the continuation parameter, in our case the capillary number, as unknown in the Newton iteration and constraining the solution along the tangent to the solution branch curve~\cite{keller1987lectures,allgower2012numerical}. The bifurcation diagram is shown in Figure~\ref{fig:bifurcationDiagram}: starting from the edge state for $\Ca=0.1$ and decreasing Ca, the droplet elongation of the equilibrium state first increases and then decreases, with a concavity developing in the central part of the droplet. When $\Ca < 0.07$, a second unstable eigenvalue appears, see for instance the eigenvalue spectra for the edge state at $\Ca=0.1$ compared to $\Ca=0.05$ which are shown in the top-right and top-left inset of figure~\ref{fig:bifurcationDiagram} respectively. These states have more than one unstable eigendirection, and are thus not edge states in the strict sense of being an attractor for the dynamics within the basin boundary. However, the states dynamically still act like edge states for two reasons: first, the second unstable eigenvalue is very small compared to the first one with the ratio being less than $10^{-2}$, second, the eigenmode associated to the second unstable eigenvalue is asymmetric and therefore it is not excited by the symmetric initial shapes hereby considered (eigenvalues associated to symmetric/non-symmetric modes are denoted by circles/crosses in the inset of figure~\ref{fig:bifurcationDiagram}). When the capillary number is increased, the edge state droplet elongation decreases. A saddle node bifurcation is encountered when $\rm{Ca}_\text{crit} = 0.1203$, as already discussed in previous studies~\cite{buckmaster1973bursting,barthes1973deformation,blawzdziewicz2002critical}. The saddle node bifurcation connects the stable solution branch (solid line) with the unstable branch of the edge states (dashed-dotted line). The bifurcation diagram shows that, for every subcritical capillary number, there exists an edge state sharing similar properties with the one found for $\Ca=0.1$. We omit edge states for very low capillary numbers since their computations become challenging due to the increasing concavity of the droplet neck. Calculation attempts at increased numerical resolution suggest that the concavity always increases as the capillary number decreases, leading us to speculate that in the limit of $\Ca=0$ the edge state may develop a cusp and correspond to two equally-sized spherical droplets in contact.

\subsection{Numerical experiment: sudden change in flow conditions}

\begin{figure}

\center
\includegraphics[width=10cm]{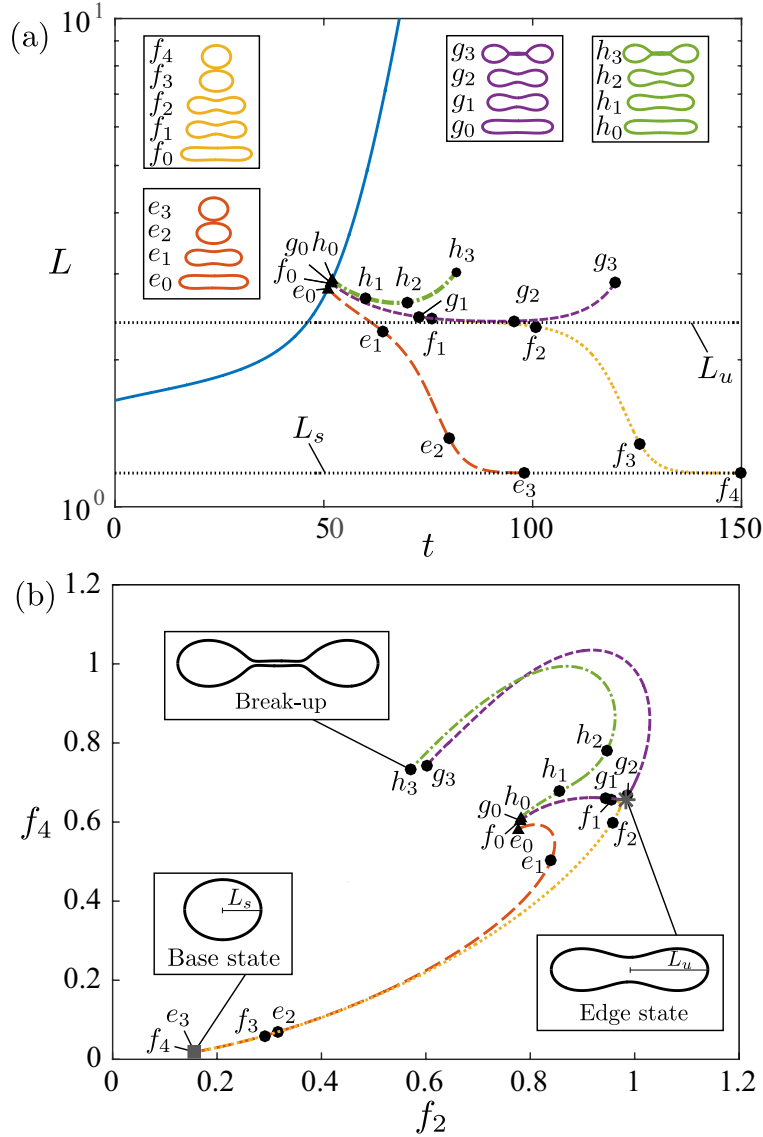}

\caption{Step change from a supercritical flow $\Ca_\text{super}=0.125$ (solid line) to a subcritical flow $\Ca_\text{sub}=\Ca_\text{super}/2$ (dashed lines), $\lambda=1$. (a) Elongation of the droplet versus time and corresponding shapes, the horizontal lines indicate the elongation of the unstable equilibrium (edge state) $L_u$ and stable equilibrium (base state) $L_s$ for $\Ca_\text{sub}=\Ca_\text{super}/2$. (b) State space orbits, the edge state selects the path towards the base state or end-pinching. The edge state and the base state are plotted and indicated respectively by the star and square symbols.}

\label{fig:likeStone}

\end{figure}

In experiments, the break-up of droplets in subcritical conditions is often a consequence of a sudden change in the flow~\cite{stone1986experimental,stone1989relaxation,stone1989influence}. For instance, an elongated droplet can result from a supercritical flow, which selects the initial condition for a subsequent subcritical flow. We address this situation by repeating the numerical experiment performed in Stone and Leal $(1989)$ and we verify the relevance of the edge state for the dynamics (see Supplemental Material video~\footnote{See Supplemental Material for the animation of the numerical experiment illustrated in figure~\ref{fig:likeStone}, from left to right: droplet elongation versus time, state space orbits and droplet shapes}). Namely, a droplet is placed in a slightly supercritical flow set by $\Ca_\text{super}=0.125$ and evolved in time with equation~\eqref{eq:time}. Since no base state exists for the selected capillary number, the droplet elongates until it breaks. However, if one applies a step change on the flow by setting $\Ca_\text{sub}=\Ca_\text{super}/2$, the droplet approaches a steady state or becomes unstable, depending on its initial elongation. This is illustrated in figure~\ref{fig:likeStone}a, with the blue solid line representing the droplet elongation for $\Ca_\text{super}$ and the dashed lines for $\Ca_\text{sub}=\Ca_\text{super}/2$ (in order of increasing initial elongation: orange long-dashed, yellow dotted, purple short-dashed, green dotted-dashed). We demonstrate the relevance of the edge state in figure~\ref{fig:likeStone}b by showing the orbits in state space. All the orbits are transiently attracted to the edge state, which selects an almost unique path towards break-up. In fact, the path toward break-up is dictated by the unstable eigenmode of the edge state, which guides the droplet elongation before final pinch off (see for instance state $h_3$ in figure~\ref{fig:likeStone}b).

The unique path thereby explains the robustness of the end-pinching mechanism for droplet break-up commonly observed in experiments. Repeating the Stone \& Leal experiment thus stresses the relevance of the edge state and also shows that its shape was indeed already observed experimentally and named the "dogbone" shape~\cite{stone1989influence}.

\begin{figure}

\center
\includegraphics[width=16cm]{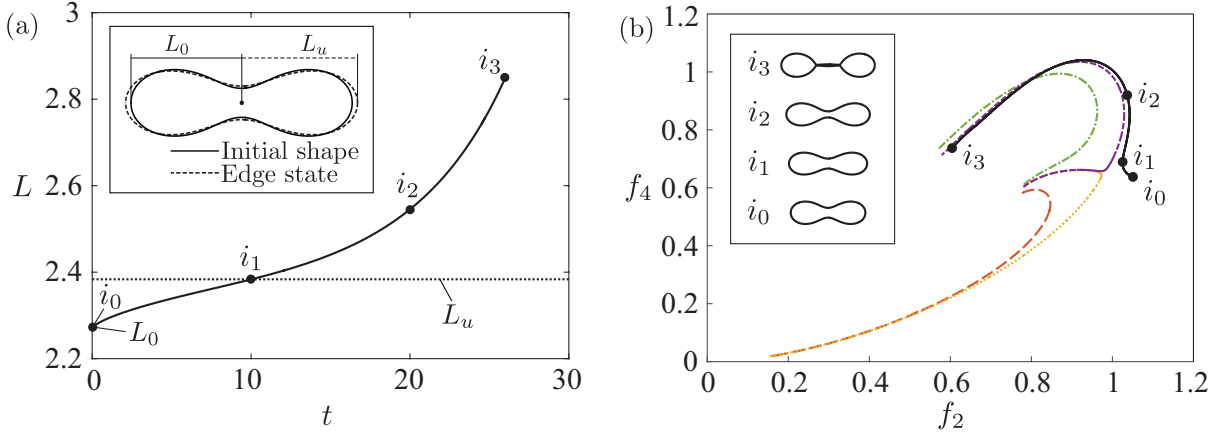}

\caption{Break-up of a droplet whose initial elongation $L_0$ is smaller than the edge state elongation $L_u$, for $\Ca=0.0625$ and $\lambda=1$. (a) Droplet elongation versus time (solid line) and edge state elongation (dotted line). The inset shows the initial shape at $t=0$ and the edge state. (b) State space orbit plotted together with the orbits in figure~\ref{fig:likeStone}. The inset shows snapshots corresponding to the markers.}

\label{fig:UnstableButLessElongated}

\end{figure}

It is worth-noting that the edge state, while it explains the robustness of the end-pinching mechanism, does not yield a critical value for the droplet elongation. In fact, the critical elongation depends on the droplet shape itself: some initial droplet shapes are stable when more elongated than the edge state (for instance $e_0$ and $f_0$ in figure~\ref{fig:likeStone}) while others become unstable when are less elongated, as reported in figure~\ref{fig:UnstableButLessElongated}a. At the same time this shows that one scalar parameter (here the droplet elongation) is not sufficient to describe the basin boundary of this high-dimensional system. However, even if these initial droplet shapes differ in term of initial elongation, they undergo a similar time evolution which is guided by the edge state and leads to the base state or break-up through the end-pinching mechanism (see figure~\ref{fig:UnstableButLessElongated}b).

\subsection{Influence of the viscosity ratio}

\begin{figure}

\center

\includegraphics[width=10cm]{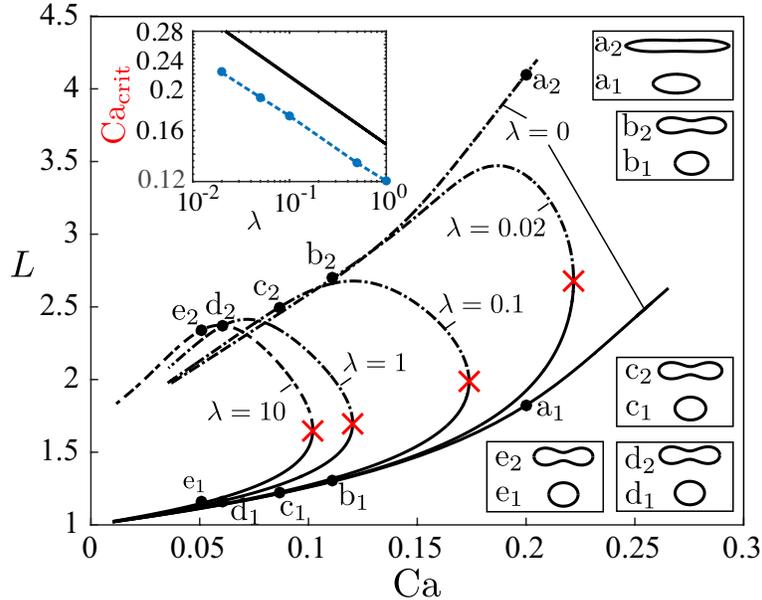}

\caption{Droplet elongation $L$  versus capillary number $\Ca$. Each line corresponds to a different viscosity ratio $\lambda$, crosses indicate the critical capillary number. Inset: log-log plot of $\Ca_\text{crit}$ versus $\lambda \in [0.02, 1]$ (dashed line) compared with the theoretical prediction of~\cite{FLM:379864} (solid line).}

\label{fig:visc}

\end{figure}

We verify that the guiding role of the edge state is robust when the viscosity ratio is varied. Continuation of nonlinear equilibrium states in the capillary number for different viscosity ratios are shown in figure~\ref{fig:visc}. The critical capillary number increases with decreasing viscosity ratio~\cite{FLM:379864,bentley1986experimental} allowing for more elongated stable base states~\cite{rallison1984deformation}. For slender droplets, the critical capillary number depends on the viscosity ratio as $\Ca_\text{crit}=0.148\lambda^{-1/6}$, based on slender body theory~\cite{FLM:379864}. We find a compatible scaling based on our fully nonlinear computations, as shown in the inset of figure~\ref{fig:visc}.
The dynamics at subcritical $\Ca$ discussed above is robust for all non-zero viscosity ratios: A finite deformation of the stable base state is required to trigger the break-up and its dynamics is controlled by the edge state, which remains connected to the base state via a saddle-node bifurcation at $\Ca_\text{crit}$. In the $\lambda \to 0$ limit corresponding to an ideal bubble with vanishing viscosity  there is no finite critical capillary number but the base state remains stable for all $\Ca$ and a finite deformation is required to trigger the break-up dynamics. Remarkably, the unstable upper branch of edge states still exists although it stays separated from the stable lower branch of base states. This separation persists at arbitrary large $\Ca$.

\section{Conclusions}

In summary, let us reiterate that at subcritical capillary numbers, the edge state equilibrium located within the basin boundary of the stable base state controls whether an initial droplet deformation leads to break-up. If the break-up is triggered, the dynamical evolution is attracted towards the edge state along its stable manifold and then repelled along the one-dimensional unstable manifold. Thereby the edge state selects an almost unique break-up path which physically corresponds to the well-known end-pinching mechanism. It is worth noting that the guiding role of the edge state is robust for varying capillary numbers and viscosity ratios. In the ideal bubble limit, there is no saddle-node bifurcation and the edge state plays a key role  in selecting the break-up mechanism for arbitrary large $\Ca$.

Our results rationalize previous studies about the mechanism responsible for droplet break-up in sub-critical extensional flows. Moreover, we answer previous conjectures regarding the existence of unstable states, finding that the "dogbone" shape, already observed in experiments, is an edge state. In our future work, we plan to complement the identification of the basin boundary by a nonlinear optimal growth analysis~\cite{pringle2010using,cherubini2011minimal}, which provides the minimal perturbation amplitude triggering break-up.

\section{Acknowledgments}

We acknowledge Mirko Farano and Florian Reetz for the helpful discussions. The European Research Council is acknowledged for funding the work through a starting grant (ERC SimCoMiCs 280117).

\bibliographystyle{ieeetr}

\end{document}